\documentstyle[12pt,epsf]{article}

\catcode`@=11
\@addtoreset{equation}{section}
\catcode`@=12

\def\sub#1{_{\rm #1}}
\def\ket#1{\left| #1 \right>}
\def\Eqn#1{Eq.~(\ref{#1})}
\def\chem#1#2#3{\begin{picture}(22,12)(0,-3)
\put(10,0){\makebox(0,0)[l]{\sf #1}} 
\put(10,5){\makebox(0,0)[r]{$\scriptstyle #2$}}
\put(10,-3){\makebox(0,0)[r]{$\scriptstyle #3$}}
\end{picture}}

\begin{document}
 
\title {\centerline{\normalsize SINP/TNP/98-03 \hfill hep-ph/9802208} 
{\bf Neutrino properties\\ from reactor and accelerator
experiments}\thanks{\sf Plenary talk given at the {\em ``B and Nu
Workshop''}\/ held at the Mehta Research Institute, Allahabad, India,
from 4 to 8 January 1998.}}
 
\author{\bf Palash B. Pal \\
\normalsize Saha Institute of Nuclear Physics, 
Block-AF, Bidhan-Nagar, Calcutta 700064, INDIA}
 
\date{}
\maketitle

\begin{abstract}
In this talk, I discuss the general theory of neutrino oscillation
experiments, putting special emphasis on the momentum distribution of
the incoming neutrino beam. Then I discuss recent
neutrino oscillation experiments, viz., LSND, KARMEN and
CHOOZ. Experiments foreseeable in the near future have also been
discussed at the end.
\end{abstract}

\section{Introduction}
	%
Static properties of any particle include its mass, charge, spin,
magnetic moment etc. For neutrinos, the charge is believed to be
zero from charge conservation, and few questions are raised about
this conclusion. There are models in which neutrinos can exhibit
a small charge, but experiments have not paid attention to
them. Similarly, the spin is known to be $1\over2$, and no one
thinks it would be interesting to challenge this value. The main
unknown property of neutrinos is their masses. For various good 
reasons, the question of mass has become the obsession of
neutrino physics in the last few decades.

Information about neutrino mass can be obtained in a variety of
laboratory experiments. We can divide these experiments into some
distinct categories. For this, we first need to mention that in the
standard model of electroweak interactions, neutrinos are taken to be
massless particles. Thus, looking for neutrino mass is to go beyond
the standard model. This can be done in two ways. First, one can study
processes which are allowed by the standard model, but try to look for
deviations in their rate from that predicted by the standard model
with massless neutrinos. These are called kinematic tests, and these
need not be accelerator or reactor experiments. For example, one can
try to find the mass of the electron neutrino from the electron
spectrum in a beta-decay process, which will not involve any
accelerator or reactor. Thus, such tests are not the topic of this
talk, although they will be mentioned at the end.

A second method would be to look for processes which are
forbidden in the standard model, but would be possible if
neutrinos have masses. Here again, we want to make two
subdivisions. One type would be processes which would require neutrino
mass but not necessarily 
neutrino mixing. Examples of such kind are neutrinoless double beta
decay or neutrino magnetic moment. Again, these are not accelerator or
reactor experiments, and so will not be discussed except at the very
end. 

Finally, there would processes which are not allowed unless there is
neutrino mixing, e.g., neutrino oscillations. These use neutrino
sources from reactors and accelerators, and we are going to describe
the theory of these experiments, and some very interesting recent
results. We will adopt a level of a person with some knowledge of
weak interaction physics, but not a great deal of knowledge of
experimental designs. This choice is not dictated by the composition
of the audience. In fact, this reflects the level of knowledge of the
speaker.

\section{Theoretical discussion of oscillation
experiments}\label{os} 
	%
We said earlier that neutrino oscillation is a kind of phenomena
which cannot occur without neutrino mixing. Let us then start by
explaining what neutrino mixing is.

Consider neutrinos created in a charged current experiment, where
either the initial state had a charge lepton $\ell$ or the final
state has, apart from the neutrino, a charged antilepton
$\hat\ell$. Let us call this neutrino $\nu_\ell$. For example, if
one thinks of an inverse $\beta$-decay process where the initial
state was $e^-+p$, the final state neutrino will be called
$\nu_e$. If one thinks of $\pi^+$ decay, if the charged particle
in the final state is $\mu^+$, the neutrino would be called
$\nu_\mu$.

This is, in fact, the sense in which so far we have used the names
$\nu_e$, $\nu_\mu$ etc. But there is a catch. No one told us that
the neutrino states defined this way would be the eigenstates of
the Hamiltonian. Suppose they are not. The eigenstates will be
some linear combinations of these states. Calling these eigenstates by
$\nu_1$, $\nu_2$ etc, we can write the linear superposition as
	\begin{eqnarray}
\left(  \begin{array}{c} \nu_e \\ \nu_\mu
\end{array}\right) =
\left(  \begin{array}{cc} \cos \theta & \sin \theta \\
         - \sin \theta & \cos \theta \end{array}\right) 
\left(  \begin{array}{c} \nu_1 \\ \nu_2
\end{array}\right) \,,
\label{os.U2}
	\end{eqnarray}
where $\nu_1$, $\nu_2$ etc are eigenstates. If this is the case, we
would say that the neutrinos are mixed, and the phenomenon would be
called neutrino mixing. More generally, for multiple neutrino states,
we should write
	\begin{eqnarray}
\nu_\ell = \sum_a U_{\ell a} \nu_a \,,
	\end{eqnarray}
where the subscript $\ell$ would run over the states $\nu_e$,
$\nu_\mu$, $\nu_\tau$ etc, and $\nu_a$'s would stand for the
eigenstates. However, for the sake of simplicity, we will work in a
two-level system.

Suppose now we have created an initial beam
	\begin{eqnarray}
\ket {\nu (0)} \equiv \ket{\nu_\alpha} 
&=& \cos \alpha \ket{\nu_e} + \sin \alpha
\ket{\nu_\mu} \nonumber\\*
&=& \cos (\theta+\alpha) \ket{\nu_1} + \sin (\theta+\alpha)
\ket{\nu_2} \,. 
	\end{eqnarray}
After this beam travels for a time $t$, the state would evolve to
	\begin{eqnarray}
\ket{\nu (t)} = e^{-iHt} \ket{\nu_\alpha} 
&=& e^{-iE_1t} \cos (\theta+\alpha) \ket{\nu_1} +
e^{-iE_2t}  \sin (\theta+\alpha) \ket{\nu_2} \nonumber\\*
&=& e^{-iE_1t} \left[ \cos (\theta+\alpha) \ket{\nu_1} +
e^{-i\Delta E\,t}  \sin (\theta+\alpha) \ket{\nu_2} \right] \,,
	\end{eqnarray}
where
	\begin{eqnarray}
\Delta E = E_2-E_1 \approx {\Delta m^2 \over 2K} \,,
	\end{eqnarray}
where $K$ is the magnitude of the 3-momentum. 
In writing the last step, we assumed that the neutrinos are
ultra-relativistic, so that the energy-momentum relation can be
approximated by
	\begin{eqnarray}
E = \sqrt {K^2 + m^2} \approx K + {m^2 \over 2K}
\,,
	\end{eqnarray}
neglecting higher powers of mass.

Now suppose we are trying to find the state $\nu_\beta$, which is
defined similar to $\nu_\alpha$ except the angle $\alpha$ replaced by
the angle $\beta$. The probability for this would be
	\begin{eqnarray}
\left| \left< \nu_\beta \right.\left| \nu(t) \right> \right|^2
&=& \left| \cos (\theta+\alpha) \cos (\theta+\beta) + 
e^{-i\Delta E\,t}  \sin (\theta+\alpha) \sin(\theta+\beta) \right|^2 
\nonumber\\*
&=& {1\over 2} \left[ 1+ \cos (2\theta+2\alpha) \cos (2\theta+2\beta) +
\sin (2\theta+2\alpha) \sin (2\theta+2\beta) \cos \left( {\Delta m^2
\over 2E} x \right) \right] \nonumber\\*
&=& {1\over 2} \left[ 1+ \cos (2\alpha - 2\beta) \right] 
- \sin (2\theta+2\alpha) \sin (2\theta+2\beta) \sin^2 \left( {\Delta m^2
\over 4K} x \right)  \,.
	\end{eqnarray}
replacing $t$ by $x$ since the neutrinos are ultra-relativistic
anyway.

In particular, if we consider $\alpha=\beta$, i.e., we 
try to find the same state that was created at $t=0$, we obtain the
``survival probability'' of that state to be
	\begin{eqnarray}
P \sub{surv} = 1 - \sin^2 (2\theta+2\alpha) 
\sin^2 \left( {\Delta m^2 \over 4K} x \right)  \,.
	\end{eqnarray}
On the other hand, if $\ket{\nu_\beta}$ is orthogonal to the original
state $\ket{\nu_\alpha}$, we obtain is the ``conversion probability'',
which is 
	\begin{eqnarray}
P \sub{conv} = \sin^2 (2\theta+2\alpha) 
\sin^2 \left( {\Delta m^2 \over 4K} x \right)  \,.
	\end{eqnarray}
If we consider only the pure flavor states, i.e.,  $\nu_e$ or
$\nu_\mu$ for the initial and final states, $\alpha$ is 
either $0$ or $\pi/2$. In both cases, we obtain
	\begin{eqnarray}
P \sub{surv} &=& 1 - \sin^2 2\theta
\sin^2 \left( {\Delta m^2 \over 4K} x \right)  \,,\nonumber\\*
P \sub{conv} &=& \sin^2 2\theta  
\sin^2 \left( {\Delta m^2 \over 4K} x \right)  \,.
	\end{eqnarray}

To connect this formula to the results of real experiments, we first
need to realize 
that in a given experiment, the neutrinos can never come with a
well-defined value of $K$ due to the uncertainty relation.
Thus, the conversion
probability obtained in a real experiment should be given by
	\begin{eqnarray}
P \sub{conv} (x) = \sin^2 2\theta \int dK \; \Phi(K) \, \sin^2
\left( {\Delta m^2 \over 4K} \, x \right) \,,
\label{os.Pconv} 
	\end{eqnarray}
where $\Phi(K)$ is the spectrum of the
incoming beam, normalized by
	\begin{eqnarray}
\int dK \; \Phi(K) = 1 \,.
\label{os.phi} 
	\end{eqnarray}

\begin{figure}
{\centerline{\epsfxsize=0.7\textwidth
\epsfysize=0.3\textheight
\epsfbox{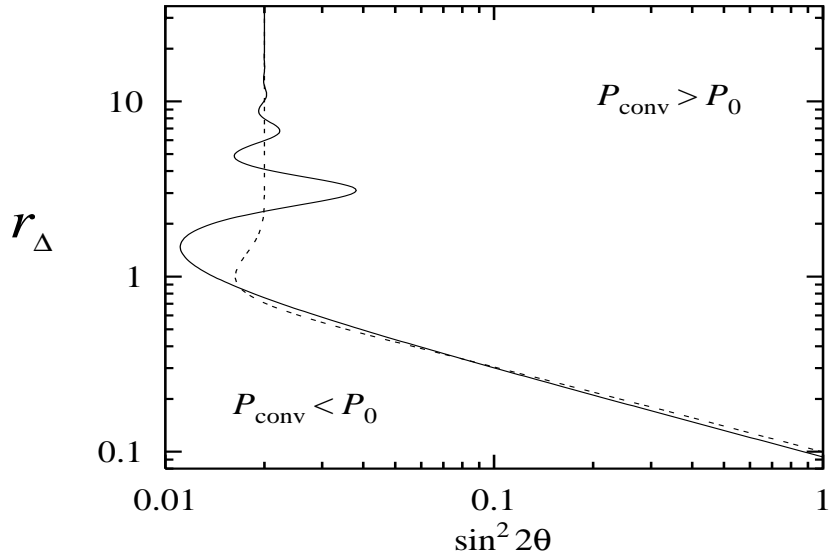}}}

\caption[]{\sf Implication of a hypothetical neutrino oscillation
experiment with a gaussian momentum distribution, with
$\sigma_K/\langle K \rangle = 0.2$, which obtains $P\sub{conv}=0.01$,
or, equivalently, $P\sub{surv}=0.99$. The dashed line represents the
approximate form of the equi-probability contour given in
\Eqn{os.approx}.} \label{os;eqP0}

\end{figure}
Suppose now in an experiment one obtains $P\sub{conv}=P_0$. What will
it mean? $\Phi(K)$ should be known from the design of the experiment
and the nature of the source. One is then left with two variables,
$\sin^22\theta$ and $\Delta m^2$. For reasons that would be clear as
we proceed, it is useful to use the dimensionless quantity
	\begin{eqnarray}
r_\Delta = {\Delta m^2 \over 4 \langle K \rangle } \, x 
	\end{eqnarray}
instead of $\Delta m^2$ directly. The conversion probability can then
be written as
	\begin{eqnarray}
P \sub{conv} (x) = \sin^2 2\theta \int dK \; \Phi(K) \, \sin^2
\left( r_\Delta {\langle K \rangle \over K} \right) \,.
\label{os.Pconv2}
	\end{eqnarray}
Thus, the result of this experiment
would pick up an allowed contour in a plot of $r_\Delta$ vs
$\sin^22\theta$. The results for $P_0=0.01$ are shown in
Fig.~\ref{os;eqP0}, using a gaussian distribution of momenta:
	\begin{eqnarray}
\Phi(K) = {1\over \sigma_K \sqrt{2\pi}} \; \exp \left( -\, {(K-
\langle K \rangle)^2 \over 2\sigma_K^2} \right) \,,
\label{os.gauss} 
	\end{eqnarray}
where we have taken $\sigma_K=0.2\langle K \rangle$ for the sake of
illustration.

Let us discuss the asymptotic features of this figure. 
If $r_\Delta$ is large, it is useful to rewrite \Eqn{os.Pconv2} in
the form
	\begin{eqnarray}
P \sub{conv} (x) = {1\over 2} \sin^2 2\theta \int dK \; \Phi(K) \,
\left[ 1 - \cos 
\left( 2r_\Delta {\langle K \rangle \over K} \right) \right] \,.
	\end{eqnarray}
The
argument of the cosine function oscillates violently
as a function of momentum in the region
where $\Phi(K)$ is appreciably different
from zero. Thus, the integral of this part vanishes. The other part
gives 
	\begin{eqnarray}
\sin^22\theta = 2P_0 \qquad \mbox{for $r_\Delta \gg 1$}\,,
\label{os.rlarge}
	\end{eqnarray}
using \Eqn{os.phi}.  This is the vertical part of the line at the top
of Fig.~\ref{os;eqP0}.

On the other hand, if $r_\Delta$ is small, 
we can replace the sine function by its argument. \Eqn{os.Pconv2} now
takes the form
	\begin{eqnarray}
P_0 &=& r_\Delta^2 \sin^2 2\theta \; \langle K \rangle^2 \int dK \;
\Phi(K) {1 \over K^2} \nonumber\\* 
&=& r_\Delta^2 \sin^2 2\theta \; \langle K \rangle^2 \left< {1 \over
K^2} \right> \,.
	\end{eqnarray}
This can be rewritten as
	\begin{eqnarray}
r_\Delta^2 \sin^2 2\theta = P_0' \qquad \mbox{for $r_\Delta \ll 1$} \,,
\label{os.rsmall}
	\end{eqnarray}
where
	\begin{eqnarray}
P_0' 
= {P_0 \over \langle K \rangle^2 \left< {1 / K^2} \right>} \,.
\label{P0'}
	\end{eqnarray}
The denominator of the right side of this equation is determined by
the momentum distribution. Therefore, $P_0'$ is known from the results
of the experiment. \Eqn{os.rsmall} now says two things. First, the
slope of the equi-probability contour in the log-log plot of
$r_\Delta$ vs $\sin^22\theta$ should be $-{1\over2}$, as seen in the
lower part of Fig.~\ref{os;eqP0}. Second, when
$\sin^22\theta=1$, the value of $r_\Delta$ for the equi-probability
contour will be given by
	\begin{eqnarray}
r_\Delta = \sqrt{P_0'} \qquad \mbox{for $\sin^22\theta=1$.}
	\end{eqnarray}
This, in fact, is the lowest value of $r_\Delta$ probed by the
experiment.

Since small $r_\Delta$ means more sensitive information about masses,
let us spend a little bit of time on this formula. Consider what
happens if we have a momentum distribution with very little spread. In
this case, we can write $\langle 1/K^2 \rangle \approx 1/\langle K
\rangle^2$, so that the denominator of \Eqn{P0'} would be unity. In
this case, $P_0' \approx P_0$, so that the lowest value of $r_\Delta$
probed by the experiment is really $\sqrt{P_0}$. This is the case with
the curve of Fig.~\ref{os;eqP0}, for which a small momentum spread
was assumed. However, in general, $P_0'\neq P_0$. Rather,
	\begin{eqnarray}
P_0' \leq P_0 
	\end{eqnarray}
since in any distribution of a non-negative variable, $\langle K
\rangle^2 \left< {1 / K^2} 
\right> \geq 1$. This is because larger values of the variable will be
preferentially picked up by the average, whereas smaller values will
be favored in the evaluation of $\left< {1/K^2} \right>$. Thus, the
spread of momentum of the initial beam is not a hassle that one has to
cope with in these experiments. On the contrary, it is helpful to have
a good spread, which allows one to probe lower and lower values of
$r_\Delta$. 

Notice that although we used a specific form of the momentum
distribution for plotting Fig.~\ref{os;eqP0}, we have never used it in
discussing the asymptotic behavior of the plot. Thus, we can say that
no matter what the momentum distribution is, we would obtain the
asymptotic forms given in \Eqn{os.rlarge} and \Eqn{os.rsmall}. Of
course, the value of $P_0'$ will depend on the distribution, and so
will the wiggles in the plot which appear in the region around
$r_\Delta\sim 1$, between the two asymptotes. If we do not care about
these wiggles, the rest of the equi-probability line can be
represented by the analytical formula
	\begin{eqnarray}
\sin^22\theta = 2P_0 + \left( {P_0' \over r_\Delta^2} - 2P_0 \right)
\exp (-r_\Delta^2) \,.
\label{os.approx}
	\end{eqnarray}
This has also been shown in Fig.~\ref{os;eqP0} with a dashed line,
taking $P_0'=P_0$ in view of the low standard deviation used for the
plot.

Of course in real experiments one does not obtain the conversion or
the survival probability without any error bar. If the experiment is a
null experiment, i.e., if they find that the conversion probability,
e.g., is less than some value, this excludes the upper right portion
of the parameter space. On the other hand, if the experiment obtains a
positive signal, we need to draw two equi-probability lines
corresponding to the smallest and the largest probabilities allowed,
and the region between two lines will be allowed.

\section{Different classes of oscillation experiments}
We now classify the existing experiments into various categories for
the clarity of presentation. The classifications will be done on
various grounds.

\paragraph*{Values of $\langle K \rangle $~:} The value of the energy
of the 
neutrinos is determined by the type of sources of the neutrinos. For
reactor neutrinos, the neutrinos are produced as by-products of
nuclear reactions, typically fission reactions, and so the energies
are of the order of a few MeV. The energies are not high enough to
produce the charged muon or tau, so no $\nu_\mu$ or $\nu_\tau$ is
available in this source. Moreover, since fission reactions produce
nuclei of lower atomic number in which the neutron to proton ratio is
usually lower, at the nucleon level these reactions convert neutrons
to protons via the reaction $n\to p+e^-+\bar\nu_e$. Hence the neutrino
source is $\bar\nu_e$ for this kind of experiments. Most of the early
experiments are of this type. Recently, the CHOOZ experiment have
given their first results. We will discuss the results later.

Another kind of experiments use medium energy neutrinos, which come in
the range of a few tens of MeVs. The sources here are typically pions
created by beam dump of a proton beam. At present, there are two
oscillation experiments which use such neutrinos. They are KARMEN and
LSND. 

High energy accelerator neutrinos, with energies in the range of
GeVs, are also beginning to be used. Many such experiments are also in
the planning or constructing stages, and we will mention some.

\paragraph*{Values of $x$~:} Early experiments on neutrino oscillation
used path lengths of the order of at most a few hundred meters.  As
indicated above, any given experiment, with a given accuracy of
detection, will probe the parameter space only upto a certain value of
$r_\Delta$ which is roughly equal to $\sqrt{P_0'}$. So, to probe lower
values of $\Delta m^2$, one needs large $x$. Thus, some long baseline
experiments are also being planned.

To obtain a feeling for the magnitudes, it is useful to use the form
	\begin{eqnarray}
r_\Delta = 1.27 \left( {\Delta m^2 \over 1\, {\rm eV}^2} \right)
\left( {\langle K \rangle  \over 1\, {\rm MeV}} \right)^{-1} 
\left( {x \over 1\, {\rm m}} \right) \,.
	\end{eqnarray}
Thus, for example, if one wants to probe $\Delta m^2$ values at the
level of $10^{-2}\,{\rm eV}^2$ with an experiment where $P_0\approx
10^{-2}$, one needs $x/\langle K \rangle \sim 10\,{\rm m/MeV}$. This
must be taken only as a benchmark value, rather than a firm estimates,
because we used $\sqrt{P_0}$ as the minimum value of $r_\Delta$ that
can be probed. As we remarked earlier, this is true only if the
standard deviation of the beam is very small compared to the mean
value. In real experiments, this is not often the case.

\paragraph*{Appearance and disappearance experiments~:} The
experiments either look for the survival probability of a certain kind
of neutrinos, or the conversion probability of a particular flavor to
another particular flavor. The first kind of experiments are called
{\em disappearance type}, and the second kind {\em appearance type}\/
of experiments.

If there were only two kinds of neutrinos, the two approaches would
provide the same information. But there are at least three kinds of
neutrinos, so this is not true. There are advantages and
disadvantages of both types. The appearance experiments are easier
in the sense that it is easier to detect a flavor which was not
present in the original beam --- even one event would signal a
non-trivial signal. In the disappearance experiments, on the other
hand, one has to check meticulously whether the flux of the original
beam has decreased over the distance.

However, appearance experiments can check only a specific channel. For
example, suppose we try to do an experiment to see whether $\nu_e$
oscillates to $\nu_\mu$, and we find a null signal to the accuracy of
the experiment. So $\nu_e$ does not oscillate to $\nu_\mu$, but who
knows, maybe it still oscillates substantially to $\nu_\tau$, or some
other unknown variety of neutrinos. If on the other hand we try a
disappearance experiment of an original $\nu_e$ beam, we would get a
non-trivial result no matter what $\nu_e$ oscillates to. But it is
also true that in this experiment, we will not know what $\nu_e$
really oscillates to.

\section{Recent results of oscillation experiments}
We now present the results of some recent experiments, and prospects
for some upcoming ones.

\subsection*{LSND~} 
The \underline Liquid \underline Scintillator \underline Neutrino
\underline Detector group, working at the Los Alamos Meson Physics
Facility (LAMPF), have consistently reported some non-null results
over the last few years. If confirmed, these will constitute the first
positive evidences for neutrino oscillations.

For the neutrino source, they produce pions by dumping 800 MeV protons
on a target. Due to charge conservation, $\pi^+$ is produced
predominantly. Then $\pi^+$ decays to $\mu^++\nu_\mu$, and
subsequently $\mu^+$ decays to $e^++\nu_e+\bar\nu_\mu$. Thus, they
have $\nu_e$, $\nu_\mu$, and $\bar\nu_\mu$ in equal amounts.

\begin{figure}
\centerline{\epsfxsize=0.7\textwidth
\epsfysize=0.4\textheight
\epsfbox{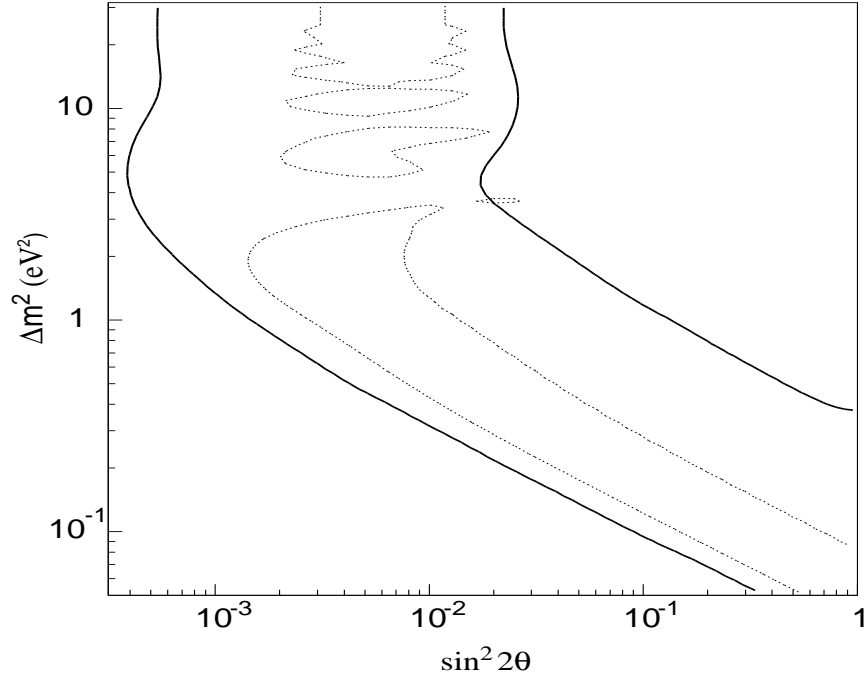}}
\caption[]{\sf Results of the two LSND 
experiments. The dashed lines indicate the allowed region from the
$\bar\nu_\mu \to \bar\nu_e$ experiment, and the solid lines the same
for the $\nu_\mu \to \nu_e$ experiment. From
Ref.~\cite{os|LSND2}.}\label{os;lsnd}
\end{figure}
The first experiment \cite{os|LSND1} of this group sought for
$\bar\nu_\mu \to \bar\nu_e$ oscillations, with the $\bar\nu_\mu$
coming from $\pi^+$ decays at rest. Recently, they have published
\cite{os|LSND2} their results for $\nu_\mu \to \nu_e$ oscillations,
using decays of pions in flight in order to obtain a broad momentum
distribution. We show both results in Fig.~\ref{os;lsnd}.

A few things can be noticed from their plots. First, since they used
in-flight pions for the $\nu_\mu \to \nu_e$ oscillations, the spectrum
was very broad, and consequently the wiggles have been almost smoothed
out. Second, note that CPT invariance tells us that the mixing angles
and $\Delta m^2$ for both these cases should be the same. Indeed, the
regions allowed by the two have a lot of intersection, which serves as
a good cross check for the results.

\begin{table}
\caption[]{\sf Details of the LSND and KARMEN
experiments.\label{os.t.2expts}} 
\begin{center}
\begin{tabular}{|p{0.2\textwidth}|p{0.3\textwidth}|p{0.3\textwidth}|}
\hline 
&& \\ 
& \multicolumn{1}{|c|}{\Large LSND} & \multicolumn{1}{|c|}{\Large
KARMEN} \\  && \\ 
\hline 
Source & Beam-dump of 800 MeV protons from linear accelerator
& Beam-dump of 800 MeV protons on a Ta-D$_2$O target  \\ \hline 
Production Mechanism & \multicolumn{2}{|c|}
{$\pi^+ \to \mu^+ + \nu_\mu$, and
subsequently $\mu^+ \to e^+ + \nu_e + \bar \nu_\mu$} \\ \hline 
Produced flavors & \multicolumn{2}{|c|} {$\nu_e$, $\nu_\mu$,
$\bar\nu_\mu$ (in equal amounts)}  \\ \hline 
Energies & \multicolumn{2}{|c|}
{\begin{tabular}[t]{l@{~:~}l} $\nu_\mu$ & 30 MeV
(mono-energetic) \\ 
$\nu_e$ and $\bar\nu_\mu$ & 0 to 53 MeV (different dispersions) \\
\end{tabular} }\\  \hline
$P_0$ for $\nu_\mu\to \nu_e$ &
$(2.6 \pm 1.0 \pm 0.5)\times 10^{-3}$ & 
$<2.0\times 10^{-2}$ \\
$P_0$ for $\bar\nu_\mu\to \bar\nu_e$ & $(3.1 ^{+1.1}_{-1.0} \pm
0.5)\times 10^{-3} $ & $<4.3\times 10^{-3}$ \\
\hline
\end{tabular}
\end{center}
\end{table}

\begin{figure}
\centerline{\epsfxsize=0.7\textwidth
\epsfysize=0.4\textheight
\epsfbox{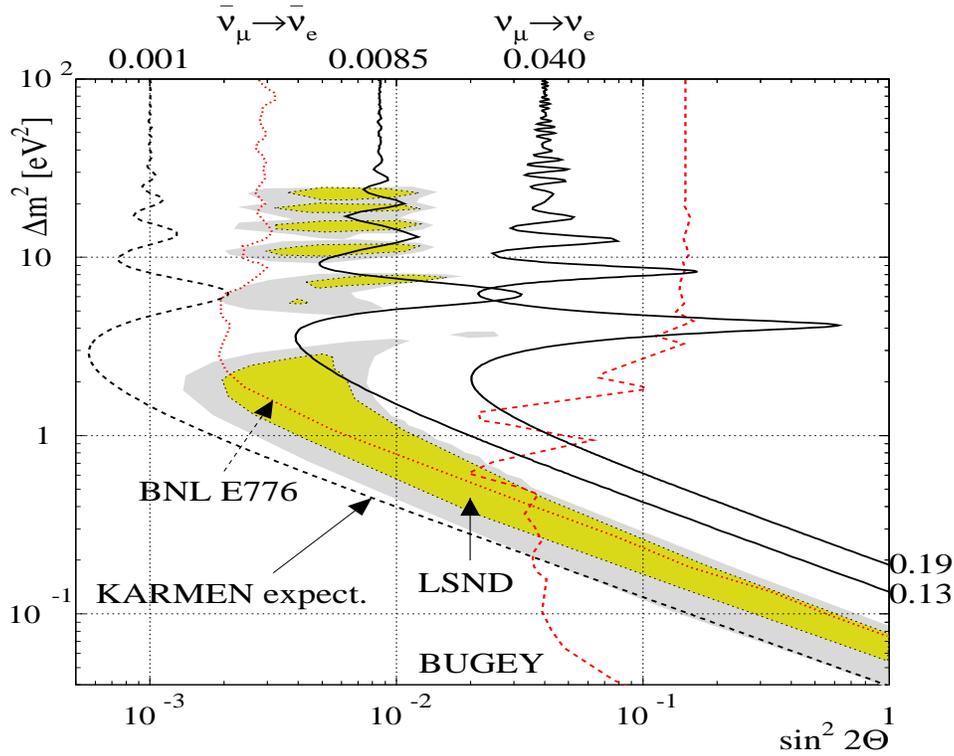}}
\caption[]{\sf Results of the KARMEN  
experiment. Also shown are the expectations for an upgraded
experiment, and the allowed regions from some earlier experiments,
including the first LSND experiment. From
Ref.~\cite{os|KARMEN}.}\label{os;karmen}
\end{figure}
\subsection*{KARMEN~} 
This is the \underline{KA}rlsruhe \underline Rutherford \underline
Medium \underline Energy \underline Neutrino experiment
\cite{os|KARMEN}. The experiment is very similar to the LSND one, in
the sense that they produce neutrinos by beam dump as well, although
the dump material is not the same. The two experiments, LSND and
KARMEN, are summarized in Table~\ref{os.t.2expts}.

The KARMEN group also, like the LSND group, have performed two kinds
of appearance experiments, $\nu_\mu\to\nu_e$ and
$\bar\nu_\mu\to\bar\nu_e$. The second one is more sensitive, since the
source has no $\bar\nu_e$. They do not find any indication of
oscillation in either experiment. They find only upper limits of
conversion probabilities. The results are shown in
Fig.~\ref{os;karmen}, where the results of some earlier experiments
are also shown, including those from a reactor experiment from the
Bugey reactor in France, and a Brookhaven experiment BNL~E776.

In the figures, one can see the region allowed by the LSND experiment
of $\bar\nu_\mu\to\bar\nu_e$. As is seen, there is no disagreement
between the two results at the moment. However, the KARMEN group is
working on an upgrade. With this, they will probe much lower values,
which are shown in the figure as well. With this upgrade, they should
see positive results if the LSND results are correct. And if they
don't see anything, that will be a contradiction with the LSND
results.

\begin{figure}
\centerline{\epsfxsize=0.7\textwidth
\epsfysize=0.4\textheight
\epsfbox{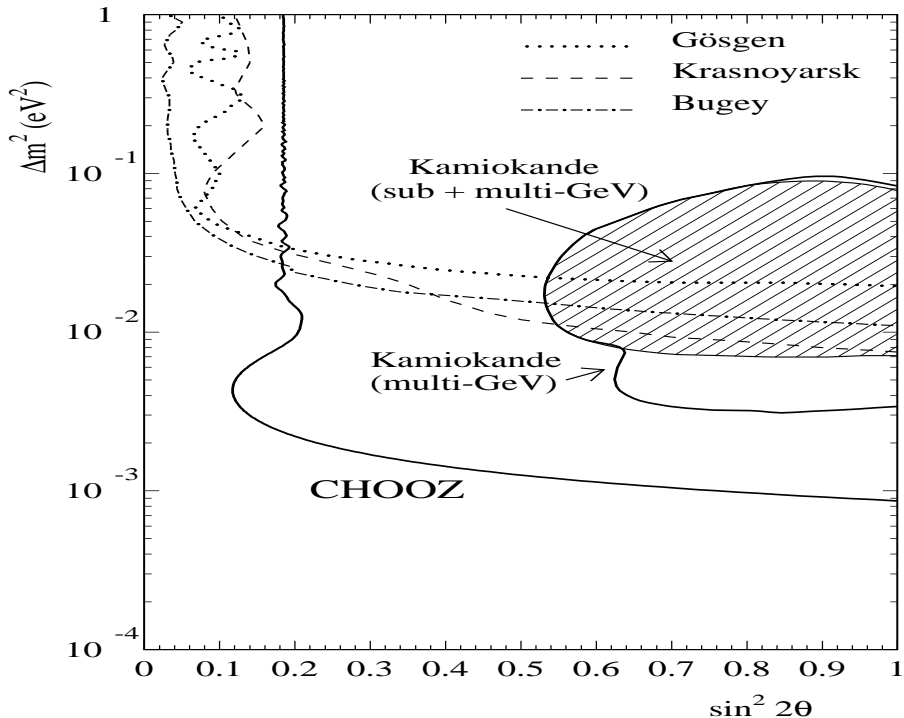}}
\caption[]{\sf Results of the CHOOZ experiment. Also shown are the
results of some earlier reactor experiments, and the region favored by
atmospheric neutrino anomaly data from Kamiokande. From
Ref.~\cite{os|CHOOZ}.}\label{os;chooz}
\end{figure}
\subsection*{CHOOZ~} 
Very recently, results have come from a reactor experiment in France
\cite{os|CHOOZ}.  The experiment is done with $\langle K \rangle
\approx 3$\,MeV and $x\approx 1$\,km. Since the source of neutrinos is
a reactor, the primary beam consists of $\bar\nu_e$. And the
experiment is of disappearance type, looking for $\bar\nu_e \to
\bar\nu_X$, where $X$ can be any flavor, including unknown ones. They
obtain
	\begin{eqnarray}
1-P_0 = 0.98 \pm 0.04 \pm 0.04 \,.
	\end{eqnarray}
This is consistent with no oscillations at all. At 90\% CL, they use
therefore 
	\begin{eqnarray}
P_0 < 0.09 \,.
	\end{eqnarray}

The important thing is that, even with a rather high value for $P_0$,
they can probe quite small values of $\Delta m^2$ since $\langle K
\rangle$ is small and the path length $x$ is large. Their results are
shown in Fig.~\ref{os;chooz}, where results of some earlier reactor
experiments have also been shown.

In addition, we have also shown the region of parameter space that can
explain the atmospheric neutrino anomaly measured by the Kamiokande
collaboration, provided the anomaly is caused by $\nu_\mu\to\nu_e$
oscillations. The CHOOZ result has ruled out the region. This of
course does not mean that either CHOOZ or Kamiokande is wrong. It
means that, if the atmospheric neutrino anomaly has to be explained by
neutrino oscillations, it must be because the $\nu_\mu$'s are being
lost by oscillating to $\nu_\tau$ or to some unknown species.

\section{Outlook for future experiments~} 
Let us reiterate the statement made earlier, that in fact an
oscillation experiment measures directly the parameter $r_\Delta$. So,
with a given sensitivity of an experiment (i.e., given $P_0$), 
one can probe smaller and smaller values of $\Delta m^2$ if one can
construct experiments with larger values of the path length $x$ and
smaller values of the average energy of the neutrino beam. It would
thus seem that if we can perform reactor experiments with path lengths
as large as possible, we would obtain the best results.

There is however a limitation to this goal. With reactor neutrinos,
first of all, one can do disappearance experiments only, and measuring
the survival probability is difficult. Secondly, reactor beams are not
very well directed. So, if we try to go to large path lengths, we also
lose considerably in the sensitivity. One can already see it in the
results of the CHOOZ experiment, which uses a path length of 1\,km
which is considerably higher than those of earlier reactor
experiments. But they paid the price by ending up with a $P_0$ which
is considerably higher compared to these earlier experiments. One can
see this from the fact that in Fig.~\ref{os;chooz}, the vertical part
of the line for large $\Delta m^2$ occurs at much higher value of
$\sin^22\theta$ for the CHOOZ experiment compared to the earlier
ones. With reactor experiments, it is hard to go much beyond in terms
of path length. Nevertheless, the Kamiokande group is contemplating 
an experiment with neutrinos from nearby reactors, where the path lengths
could well be larger than a kilometer.

Accelerator beams are however well focused and one can hope to go to
hundreds of kilometers of path length with them.  Various long
baseline experiments are in the planning or construction stage at the
moment with accelerator neutrinos. We give a brief summary of
them~\cite{os|Zuber}.

\subparagraph*{K2K~:} The distance between KEK laboratory and the
Kamiokande experiment in Japan is about 235\,km. There is a plan of
sending a neutrino beam from KEK to Kamiokande, with an average energy
of 1\,GeV. At Kamiokande, they will be detected. The beam-line should
be finished by 1998, and data should be expected by 1999.

\subparagraph*{Fermilab--Soudan~:} The distance here is 735\,km. The
MINOS experiment will be installed in the Soudan mines. The project
may start at the beginning of the next century.

\subparagraph*{CERN--Gran-Sasso~:} The distance is 732\,km. Neutrino
beam from CERN can be detected at Gran Sasso. Various proposals exists
for detection of the neutrinos at Gran Sasso.

\section{Conclusion}
Since I have started with an ``Introduction'', I have to end with a
``Conclusion''. There are of course some obvious concluding
remarks. With more experiments, we will know things better. The
important point is that at the moment, there seems to be a lot of
activity in the field. With the KARMEN upgrade, the LSND results will
be checked, and we are looking forward to it. The long baseline
experiments will usher a new era in this field, although the results
of these experiments are not expected until a few years from now.

One of the things that I want to emphasize here is that the neutrino
oscillation experiments at present provide the most sensitive
information about neutrino mass from all terrestrial experiments. Let
us compare them briefly with other kinds of experiments described in
the Introduction. The kinematic tests provide only upper bounds of
neutrino masses, which is of the order of a few eV for $\nu_e$,
170\,keV for $\nu_\mu$, and 24\,MeV for $\nu_\tau$.

Neutrinoless double beta decay process is not allowed unless lepton
number symmetry is violated. The amplitude of this process depends
on the factor
	\begin{eqnarray}
m \sub{eff} = \sum_a U_{ea}^2 m_a \,,
	\end{eqnarray}
where $m_a$ is the mass of the $a$-th neutrino eigenstate. In absence
of neutrino mixing, the quantity $m\sub{eff}$ becomes equal to the
mass of the $\nu_e$. The 
most stringent bounds come from the Heidelberg-Moscow
experiment \cite{bb|HeiMos}.
They use 19.2\,kg of $\chem {Ge}{76}{}$ and search
for the process
	\begin{eqnarray}
\chem {Ge}{76}{} \to \chem {Se}{76}{} + 2e^- \,,
	\end{eqnarray}
and set the following lower limit on the lifetime:
	\begin{eqnarray}
T_{1/2} > 7.4 \times 10^{24} \; {\rm yr}.
	\end{eqnarray}
It is not straight forward to deduce the upper bound on $m\sub{eff}$
from this raw experimental number. The reason is that the calculation
of this rate involves nuclear matrix elements, the evaluation of which
cannot be done in a manner that pleases everyone. Using various
evaluation of the matrix elements, one obtains the upper bound on
$m\sub{eff}$ ranging between $0.56$\,eV to $1.76$\,eV. One can say
that the upper limit is of the order of an eV.

Neutrino oscillation experiments, on the other hand, cannot probe the
mass values directly. They can measure the mass squared
differences. If, inspired by the mass patterns of the charged
fermions, one assumes a hierarchy in the neutrino masses, the lowest
mass value probed would be the square root of the lowest $\Delta m^2$
probed. With this assumption, we see the neutrino oscillation
experiments are already probing mass values which are lower than an
eV. With upgrades and improvements, they can go to even lower
values. Fortunately, these upgrades and improvements are not day
dreaming, they are things to expect in the near future.

\paragraph*{Acknowledgments~:} I thank the organizers of the conference
for giving me a chance to talk, and Lincoln Wolfenstein and Sandip
Pakvasa for discussions.

\end{document}